\input harvmac

\def\K3{{\bf K3}}
\def\journal#1&#2(#3){\unskip, \sl #1\ \bf #2 \rm(19#3) }
\def\andjournal#1&#2(#3){\sl #1~\bf #2 \rm (19#3) }

\def\bar{\overline}

\def\tilde{\widetilde}
\def\sst{\scriptscriptstyle}

\def\frac#1#2{{#1\over#2}}

\def\half{\frac12}

\def\inbar{\,\vrule height1.5ex width.4pt depth0pt}
\def\IC{\relax\hbox{$\inbar\kern-.3em{\rm C}$}}
\def\IR{\relax{\rm I\kern-.18em R}}
\def\IP{\relax{\rm I\kern-.18em P}}

%
%

%
\catcode`\@=11
\def\slash#1{\mathord{\mathpalette\c@ncel{#1}}}
\overfullrule=0pt

\def\underrel#1\over#2{\mathrel{\mathop{\kern\z@#1}\limits_{#2}}}

\catcode`\@=12


%

\def\det{{\rm det}}

\def\det{{\rm det}}



\rightline{IASSNS-HEP-00/33}
\rightline{SU-ITP 00-13}
\Title{
\rightline{hep-th/0005040}}
{\vbox{\centerline{Strings in Background Electric Field,}
\centerline{Space/Time Noncommutativity and}
\centerline{A New Noncritical String Theory}}}
\bigskip

\centerline{\it N. Seiberg,}
\medskip
\centerline{School of Natural Sciences}
\centerline{Institute for Advanced Study}
\centerline{Olden Lane, Princeton, NJ 08540}
\bigskip

\centerline{\it L. Susskind and N. Toumbas}
\medskip
\centerline{Department of Physics}
\centerline{Stanford University}
\centerline{Stanford, CA 94305-4060}
\bigskip

\vglue .3cm

\bigskip
\noindent
Searching for space/time noncommutativity we reconsider open strings in
a constant background electric field.  The main difference between
this situation and its magnetic counterpart is that here there is a
critical electric field beyond which the theory does not make sense.
We show that this critical field prevents us from finding a limit in
which the theory becomes a field theory on a noncommutative spacetime.
However, an appropriate limit toward the critical field leads to a
novel noncritical string theory on a noncommutative spacetime.

\Date{5/00}

\newsec{Introduction}

\nref\cds{A. Connes, M. R. Douglas, and A. Schwarz, ``Noncommutative
Geometry and Matrix Theory: Compactification On Tori,'' JHEP {\bf
9802} (1998) 003, [hep-th/9711162].}%
\nref\dh{M. R. Douglas and C. Hull, ``$D$-Branes And The
Noncommutative Torus,'' JHEP {\bf 9802} (1998) 008,
[hep-th/9711165].}%
\nref\swnoncom{N.~Seiberg and E.~Witten,
``String theory and noncommutative geometry,''
JHEP {\bf 9909} (1999) 032, [hep-th/9908142].}%

The recent interest in noncommutative geometry and its appearance in
string theory \refs{\cds-\swnoncom}, has been limited almost
exclusively to the case where the noncommutative coordinates are
spacelike:
\eqn\basiccoms{[x^i,x^j]=i\theta^{ij}.}
The time direction, $x^0$, remained unaffected and the commutator
\eqn\basiccom{[x^0,x^i]=i\theta^{0i}}
was set to zero.

It is interesting to ask what happens, if $\theta^{0i}\not=0$.  The
meaning of \basiccom\ in this case is not completely clear.  In
quantum mechanics the time coordinate labels the evolution of the
system.  Unlike the coordinates of particles, the time is not an
operator, and hence, it is not clear how it could fail to commute.
Nevertheless, there are several motivations for studying such a
situation:

\item{1.} Equation \basiccom\ with nonzero $\theta^{0i}$ is a natural
extension of \basiccoms, and as such it appears like an interesting
situation to explore.
\item{2.} Noncommutativity in space \basiccoms\ occurs when D-branes
are placed in a nonzero $B$ field with components along the
space-space directions.  Equivalently, it occurs when the D-branes
have background magnetic field.  Similarly, $\theta^{0i}$ can be
activated by turning on $B_{0i}$ or by placing the D-branes in
background electric field.  Therefore, these are physical situations
which can occur and as such should be understood.
\item{3.}  The commutator \basiccom\ leads to an uncertainty relation
between time and space of the form
\eqn\spacetimeuncert{\Delta x^0 \Delta x^i \not= 0.}
Such an uncertainty relation was advocated as a generic property of
string theory -- ``the stringy uncertainty principle'' even when no
electric field is present (see
\ref\yoneya{T. Yoneya, ``String Theory and Space-Time Uncertainty
Principle,'' [hep-th/0004074].}
for a review and an extensive list of earlier references).  In
\ref\sst{N. Seiberg, L. Susskind and N. Toumbas, 
``The Teleological Behavior of Rigid Regge Rods''
hep-th/0005015.}
we explore some of the consequences of such an uncertainty principle.
\item{4.} The stringy uncertainty principle \spacetimeuncert\ and the
space/time commutator \basiccom\ play a crucial role in the
proposal of
\ref\blackhole{L. Susskind, ``Strings,
Black Holes and Lorentz Contractions,'' Phys. Rev. {\bf D49} (1994)
6606; G. 't Hooft, ``Horizon Operator Approach to Black Hole
Quantization,'' gr-qc/9402037;  ``The scattering matrix
approach for the quantum black hole, an overview,'' gr-qc/9607022,
Int. J. Mod. Phys. {\bf A11} (1996) 4623.}
regarding the information puzzle in black-holes.
\item{5.} As we said above, lack of commutativity of time appears to
be in serious conflict with our current understanding of quantum
mechanics.  Better understanding of this issue can perhaps shed light
on the role of time in string theory.

Gauge transformation on a noncommutative space are different than on a
commutative space.  However, as in \swnoncom, there exists a change of
variables from noncommutative to commutative gauge fields allowing us
to express the effective Lagrangian in terms of ordinary commutative
gauge fields.  This transformation applies to the space/time as well
as to the space/space noncommutativity and it resolves some of the
mysteries associated with such gauge fields.  The effective Lagrangian
after the transformation still includes an infinite number of space
and time derivatives and questions of unitarity and causality might
still be present.  However, in the specific example of string theory
in background electric field the first quantized description makes it
clear that, at least in perturbation theory, the theory is unitary and
causal.

Motivated by these considerations we will study strings in background
electric field.  Our main concern will be the question of whether
there exists a limit in which the full string theory simplifies.
One's first hope is that as for magnetic backgrounds \swnoncom, there
exists an $\alpha' \rightarrow 0$ limit in which the theory becomes
field theoretic and hence more tractable.  One of our conclusions is
that this is not the case.  The best we could find is a limit of the
theory in which the closed strings including the gravitational degrees
of freedom decouple and we are left with a simpler theory of open
strings only.  However, unlike the magnetic case, this theory is not a
field theory but a string theory.

\bigskip
\centerline{\it The setup}

We will be studying D-branes in flat $\CR^n$ with metric $g_{\mu\nu}$
and constant $B$ field $B_{\mu\nu}$.  We can pick a frame in which
$g_{\mu\nu}= \eta_{\mu\nu}$, but since we are planning to scale the
metric we do not do that.  For simplicity we will take the metric
$g_{\mu\nu}$ to be diagonal.  Then it is easy to distinguish between
the spatial and the temporal directions.

The boundary conditions for open strings are
\eqn\boundcond{\left(g_{\mu\nu}\partial_\sigma x^\nu+2\pi
\alpha'B_{\mu\nu}\partial_ \tau x^\nu\right) \big|_{\rm boundary}=0,}
where $\tau$ is tangent to the boundary and $\sigma$ is normal to the
boundary.  The analysis of \swnoncom, which was motivated by
\ref\schomerus{V. Schomerus, ``$D$-Branes And Deformation Quantization,''
JHEP {\bf 9906} (1999) 030, [hep-th/9903205].},
leads to the definition of the open string metric $G_{\mu\nu}$, the
noncommutativity parameter $\theta^{\mu\nu}$ and the open string
coupling constant $G_s$
\eqn\generalopen{\eqalign{
&G_{\mu\nu}=g_{\mu\nu}-(2\pi\alpha')^2(Bg^{-1}B)_{\mu\nu}\cr
&\theta^{\mu\nu}=2\pi \alpha'\left({1\over g+2\pi
\alpha'B}\right)_A^{\mu\nu} \cr
&G_s=g_s\left({\det G_{\mu\nu}\over \det (g_{\mu\nu}+2\pi
\alpha'B_{\mu\nu})}\right)^\half,}}
where $()_A$ denotes the antisymmetric part and $g_s$ is the closed
string coupling.

As is well known the backgound $B_{\mu\nu}$ can be traded for
background $F_{\mu\nu}$ on the branes.  Most of the analysis in the
literature is devoted to the magnetic case, where only the spatial
components of $B$ (or $F$) are nonzero.  Here we will be interested in
the purely electric case with $B_{ij}=F_{ij}=0$, but $B_{0i}\not=0$.
The generalization to the case of both electric and magnetic
background fields is straightforward.

Without loss of generality we can choose a frame in which the electric
field points in the $x^1$ direction.  Focusing on $x^{0,1}$ we take
\eqn\choicese{g_{\mu\nu}=g\pmatrix{-1&0\cr 0&1\cr}, \qquad
B_{\mu\nu}=\pmatrix{0&E\cr -E& 0}.}
We also find it convenient to define the dimensionless electric field
\eqn\tildee{\tilde E= {2\pi\alpha' E \over g}={E\over E_{cr}},}
where $E_{cr} ={g \over 2\pi \alpha'}$ is the critical electric field.
In this frame the open string parameters \generalopen\ become
\eqn\openp{\eqalign{
&G_{\mu\nu}=g_{\mu\nu}-(2\pi\alpha')^2(Bg^{-1}B)_{\mu\nu}=
{g^2-(2\pi\alpha'E)^2
\over g} \pmatrix{-1&0\cr 0& 1}_{\mu\nu} \equiv G  \pmatrix{-1&0\cr 0&
1}_{\mu\nu} \cr
&\theta^{\mu\nu}={(2\pi \alpha')^2 E \over g^2 - (2\pi\alpha'E)^2}
\pmatrix{0&1\cr -1&0\cr}^{\mu\nu} \equiv \theta \pmatrix{0&1\cr
-1&0\cr}^{\mu\nu} \cr
&G_s=g_s\left({\det G_{\mu\nu}\over \det (g_{\mu\nu}+2\pi
\alpha'B_{\mu\nu})}\right)^\half =
g_s\left(1-\left({2\pi\alpha'E\over g}\right)^2\right)^\half. \cr}}
The constants $G$ and $\theta$ can be expressed as
\eqn\openpara{\eqalign{
&G= g(1-\tilde E^2)\cr
&\theta={1\over E_{cr}} { \tilde E \over 1-\tilde E^2}\cr
&G_s=g_s\left(1-\tilde E^2\right)^\half. \cr} }
In the case of $N$ branes the effective 't Hooft coupling constant is
\eqn\thooftco{G_{eff}=NG_s=Ng_s\left(1-\tilde E^2\right)^\half.}

In section 2 we study strings in background electric field.  We first
review and extend the canonical quantization of the first quantized
theory on the strip, and then discuss the interactions.  We end with a
few comments about the phenomenon of critical electric field.  In
section 3 we show that there is no zero slope limit in which the
theory becomes a field theory on a noncommutative spacetime.  Our
novel noncritical string theory on such a spacetime is presented in
section 4.

As we were completing this work we learned of a closely related work
by R. Gopakumar, J. Maldacena, S. Minwalla and A. Strominger
\ref\gmms{R. Gopakumar, J. Maldacena, S. Minwalla and A. Strominger,
to appear.}. 

\newsec{Strings In Background Electric Field}

\subsec{Canonical quantization on the strip}

\nref\burgess{C.~P.~Burgess,
``Open String Instability In Background Electric Fields,''
Nucl.\ Phys.\  {\bf B294} (1987) 427.}%
\nref\chuho{C.~Chu and P.~Ho, ``Noncommutative open string and
D-brane,'' Nucl.\ Phys.\  {\bf B550} (1999) 151, hep-th/9812219.}%

In this subsection we follow \refs{\burgess,\chuho} and analyze the
spectrum of the first quantized open string in the background of
constant electric field.

In terms of the spacetime lightcone coordinates $x^\pm = {1 \over
\sqrt 2}(x^0\pm x^1)$ the boundary conditions \boundcond\ become
\eqn\boundconde{\left(\partial_\sigma x^\pm \mp \tilde E
\partial_\tau x^\pm \right)\big|_{\sigma=0,\pi}=0.}
The solution of the worldsheet equations of motion subject to these
boundary conditions are
\eqn\normmode{\eqalign{
x^\pm(\sigma,\tau)&=x^\pm_0+\alpha'p^\pm(\tau\pm \tilde E  \sigma) -i
(\alpha')^\half \sum_{n\not= 0}{a^\pm_n \over n} \left[ \left(1 \pm
\tilde E \right) e^{in(\sigma+\tau)} + \left(1 \mp \tilde E
\right) e^{in(-\sigma+\tau)} \right] \cr
&=x^\pm_0+\half\alpha'p^\pm\left[(\tau+\sigma)(1\pm \tilde E)+(\tau-
\sigma)(1\mp \tilde E)\right] \cr
&\qquad-i (\alpha')^\half \sum_{n\not= 0}{a^\pm_n \over n} \left[
\left(1 \pm \tilde E \right) e^{in(\sigma+\tau)} + \left(1 \mp \tilde
E \right) e^{in(-\sigma+\tau)} \right],}}
with reality conditions $(a^\pm_n)^*=a^\pm_{-n}$.

We note that as $E$ approaches the critical field $\pm E_{cr}$
($\tilde E \rightarrow \pm 1$), the modes of $x^\pm$ become chiral in
the worldsheet sense.

As in \chuho, the commutation relations of the modes in \normmode\
are
\eqn\commutat{\eqalign{
&[x_0^\mu,x_0^\nu]=i\theta^{\mu\nu},\cr
&[x_0^\mu,p^\nu]=i2G^{\mu\nu},\cr
&[a^\mu_n,a^\nu_m]={1\over 2}n\delta_{n+m,0}G^{\mu\nu},}}
and the worldsheet Hamiltonian is given by
\eqn\hamiltonian{L_0={1 \over 4} \alpha' G_{\mu\nu} p^\mu p^\nu
+2G_{\mu\nu}\sum_{n=1}^\infty a^\mu_{-n}a^\nu_n +L_0^{tr},}
where $L_0^{tr}$ is the contribution of the transverse directions.
This Hamiltonian determines the masses of the various string modes.
It is important that the relevant metric in all these expressions is
the open string metric $G_{\mu\nu}$ of \generalopen\openp.

\nref\shejabba{M.~M.~Sheikh-Jabbari,
``Open strings in a B-field background as electric dipoles,''
Phys.\ Lett.\  {\bf B455} (1999) 129
[hep-th/9901080].}%
\nref\bigsus{D.~Bigatti and L.~Susskind,
``Magnetic fields, branes and noncommutative geometry,''
hep-th/9908056.}%
\nref\yin{Z.~Yin, ``A note on space noncommutativity,''
Phys.\ Lett.\  {\bf B466} (1999) 234, [hep-th/9908152].}%

In order to gain some intuition to the physics of the problem we
neglect the oscillators in \normmode
\eqn\xzeromo{\bar x^\pm(\sigma,\tau)=x_0^\pm +\alpha'p^\pm(\tau\pm
\tilde E\sigma).}
As in \refs{\shejabba-\yin} (see also \swnoncom), in this
approximation the string is a rod which behaves like a dipole -- the
two ends of the string carry opposite charges.  Two important
properties of $\bar x$ are the following
\item{1.} It is easy to check using \commutat\ that its center of mass
coordinates, $\tilde x_0^\pm=x_0^\pm\pm {\pi\over 2}\alpha'p^\pm
\tilde E$,
commute.  The lack of commutativity of $x_0^\mu$ in \commutat\ occurs
only because these are the boundaries of the string.
\item{2.} The length of the rod is proportional to $E p$.  As in the
magnetic case, the size of the dipole is proportional to the
background field and the momentum \refs{\shejabba-\yin}.  However,
here, the dipole grows in the longitudial direction rather than in the
transverse direction.  This fact plays a crucial role in \sst.

In order to find the physical spectrum of the string we should
remember to include the ghosts and look for the BRST cohomology.
Instead, we can try to use a physical gauge like a lightcone gauge.
The boundary conditions \boundconde\ do not allow us to choose
$x^+=\alpha'p^+\tau$.  Instead, a natural guess for the
lightcone gauge in this system is
\eqn\lightconeg{x^+=\alpha'p^+(\tau+\tilde E\sigma).}
The choice \lightconeg\ is natural because it satisfies the equation
of motion and the boundary conditions.  In order to prove
it, we replace the previous choice of the conformal
gauge with the metric $\pmatrix{-1 & \cr & 1}$ by a gauge choice with
the worldsheet metric $\pmatrix{-1& \tilde E \cr \tilde E & 1-\tilde
E^2 }$.  Then we can choose the standard lightcone gauge
$x^+=\alpha'p^+\tau$.  Finally, we change the worldsheet coordinates
to a frame with a metric $\pmatrix{-1 & \cr & 1}$ to find \lightconeg.

In the lightcone gauge \lightconeg\ the oscillators in $a^\pm$ are
absent and the physical spectrum of the theory is essentially that of the
transverse directions without the electric field.  The only effect
of the electric field is to change the metric $g_{\mu\nu}$ to the open
string metric $G_{\mu\nu}$ as in \generalopen.  This metric appears in
the mass shell condition through the Hamiltonian \hamiltonian.

Examining \lightconeg\ (or \xzeromo) we see that for fixed $\tau $ the
two ends of the string are not at the same value of $x^+$.  The two
ends of the string are always spacelike separated.  As the electric
field becomes critical $\tilde E=\pm 1$ the two end points at fixed
$\tau$ become lightlike separated.  This gives another perspective to
the impossibility of having larger values of the electric field.

This picture leads to another interpretation to the fuzziness in $x^+$
corresponding to the noncommutativity in it.  This is important
because in the lightcone gauge $x^+$ is treated as a parameter,
labeling the evolution of the system and therefore there is
superficially no room for any fuzziness in it.  The fuzziness occurs
because the string is not point-like but is like a rod whose two end
points are at different times.  This is similar to the origin of the
fuzziness of space in the magnetic case \bigsus\ which originates from
the size of the dipole.

An important difference between the electric and the magnetic
backgrounds is the following.  For magnetic backgrounds the dipole
grows in the direction transverse to the motion by amount proportional
to $B_{ij}p^i$, while for electric backgrounds the growth is in the
longitudinal direction by amount proportional to $B_{0i}p^i$.

Finally we would like to resolve a little puzzle.  The discussion of
the noncommutative field theory in \bigsus\ derives the size of the
dipole as $\Delta x^\mu= \theta^{\mu\nu}p_\nu$, while from the normal
mode expansion of the zero mode, as in \xzeromo, the size is $\Delta
x^\mu= -(2\pi \alpha')^2 g^{\mu\nu}B_{\nu\rho}p^\rho$.  How can these
two expressions be the same?  The answer is that in the noncommutative
field theory we use the open string metric $G_{\mu\nu}$ to raise and
lower indices.  Using the identity $\theta^{\mu\nu}G_{\nu\rho}=
-(2\pi\alpha')^2 g^{\mu\nu} B_{\nu \rho}$ (which follows from
\generalopen) and $p_\alpha=G_{\alpha\beta}p^\beta$ it is easy to show
that these two expressions for the size of the dipole are identical.

\subsec{String interactions}

The discussion so far was limited to the free first quantized theory
and its spectrum on the strip.  In order to include interactions using
vertex operators we must use a covariant gauge.  Otherwise we cannot
change the value of $p^+$, and we cannot probe the noncommutative
phases $p_1^+p_2^-\theta$ which are characteristic of the noncommutative
theory.

At the level of the disc amplitudes the inclusion of the electric
field is straightforward.  As explained in \swnoncom, all we need to
do is to start with the corresponding amplitudes for $E=0$ and replace
$g_{\mu\nu}$ by $G_{\mu\nu}$, $g_s$ by $G_s$ and multiply the
answer by the standard phase factor with the noncommutativity
parameter $\theta$.  The fact that $g$ is replaced with $G$ is
consistent with the replacement of $g^{\mu\nu}p_\mu p_\nu$ by
$G^{\mu\nu}p_\mu p_\nu$ in the mass formula (the worldsheet
Hamiltonian \hamiltonian).  These observations are consistent with the
explicit calculations performed in the special case of scattering on
D1-branes in
\ref\gkp{S.~Gukov, I.~R.~Klebanov and A.~M.~Polyakov,
``Dynamics of (n,1) strings,'' Phys.\ Lett.\  {\bf B423} (1998) 64,
[hep-th/9711112].}.

\subsec{The critical electric field}

At $E=E_{cr}$ the various expressions for the open string parameters
\openpara\ are singular.  The origin of this singularity was
understood in \burgess.  An open string should be thought of as a
dipole whose ends carry opposite charges.  When an open string is
stretched along the $x^1$ direction in the orientation preferred by
the background electric field, the field reduces its energy.  For
$|E|\roughly< E_{cr}$, the energy stored in the tension of the string
is almost balanced by the electric energy of the stretched string.
More quantitatively, the effective tension of such a string is
\eqn\effecten{{1\over 2\pi\alpha'_{eff}}= {1\over
2\pi\alpha'} {G_{11}\over g}={1-\tilde E^2\over 2\pi\alpha'},}
and it vanishes as $|E| \rightarrow E_{cr}$.  Beyond that value
virtual strings can materialize out of the vacuum, stretch to infinity
and destabilize the vacuum.

When the string is streched along different directions this effect is
absent and its tension is not reduced.  Instead of thinking of a
varied tension, it is easier to keep the tension fixed at $1\over 2\pi
\alpha'$ and scale the length.  Equivalently, we scale the
metric.
This is the intuitive reason for the expression for the open string
metric $G_{\mu\nu}$ which has the effect of reducing the metric in the
directions of $x^0$ and $x^1$ but keeping $G_{\mu\nu}=g_{\mu\nu}$ in
the other directions.

The expression for the effective open string coupling constant
\openpara
\eqn\opencoupling{G_s=g_s\left(1-\tilde E^2\right)^\half}
is also interesting.  We note that it vanishes as $E$ approaches the
critical field for fixed value of the underlying string coupling
constant $g_s$.  This fact has a simple reason.  Consider the
transition of one of the light open strings to a closed string.  For
this process to take place, the stretched open string needs to bend
over in order for its ends to touch each other.  Then, part of it will
be stretched against the electric field and will be very heavy.
Therefore, such a transition cannot happen easily and correspondingly
the effective coupling constant is small.

Finally, let us compactify $x^1$ and use T-duality along the compact
direction. We get branes of one lower dimension moving with velocity
$v = \tilde E$ around the circle. As the electric field approaches its
critical value, the velocity of the branes approaches the speed of
light and the momentum of the branes becomes large. We can then make a
large boost to bring the branes to rest. In the T-dual picture of the
original theory, this amounts to boosting to large momentum.
Therefore, studying the electric theory near the critical limit is
equivalent to studying the theory in the DLCQ description.

Below we will study the limit $E\rightarrow E_{cr}$ in more detail.
We will define an appropriate scaling limit which focuses on these
light strings and find an effective theory of these modes only.

\newsec{Looking For A Zero Slope Limit With Noncommutative Spacetime}

In this section we will look for a field theory limit of string theory
in electric field background, which has a noncommutative spacetime.
Naively, all we have to do is repeat the analysis of \swnoncom\ to
find a zero slope limit.  However, we will show that this is not
possible.

In our discussion we will hold the coordinates and external momenta
fixed, and will rescale the metric and $\alpha'$.  An alternate
description in terms of fixed metric and $\alpha'$ but rescaled
coordinates and momenta is, of course, also possible.

We start by reviewing the propagator along the boundary of the disc
\eqn\propa{\langle x^\mu(\tau) x^\nu(0) \rangle = -\alpha' G^{\mu\nu}
\log \tau^2 + i {\theta^{\mu\nu}\over 2} \epsilon(\tau),}
where the coefficients of the two terms in the propagator are given in
terms of
\eqn\openparaa{\eqalign{
&\theta={E\over E_{cr}^2-E^2}={1 \over E_{cr}} { \tilde E \over 1 -
\tilde E^2}\cr
&\alpha'G^{-1}= {1 \over 2\pi E_{cr} (1-\tilde E^2)} = {1 \over 2\pi}
{ \theta \over \tilde E} .}}

We are looking for a limit subject to the following requirements:
\item{1.} We want $\alpha'\rightarrow 0$, so that all the oscillator
modes decouple and the resulting theory is field theoretic.
Otherwise, the theory has an infinite number of massive string modes.
\item{2.} We should keep $G$ finite.  The reason for that is that
$G_{\mu\nu}$ will be the metric of the resulting theory, which should
be finite.
\item{3.} The noncommutativity parameter $\theta$ should be finite,
since we want the resulting theory to be noncommutative.

Since $|\tilde E|<1$, if $\theta$ is kept fixed, $\alpha' G^{-1}$
cannot be taken to zero.  Therefore, the first term in the propagator
\propa\ cannot be neglected and the directions $x^{0,1}$ remain
stringy.  This is unlike the situation for magnetic backgrounds,
where there is no bound on the background magnetic field, and it is
possible to take the analog of $\tilde E$ to infinity and remove the
first term in the propagator, thus making the theory in these
directions field theoretic \swnoncom.

Roughly, for magnetic backgrounds the scaling is such that the
magnetic field goes to infinity.  Here, we cannot scale $E$ to
infinity, and therefore we cannot reach a zero slope limit.

One might argue that perhaps we should scale $\alpha' \rightarrow 0$,
while also scaling $G\rightarrow 0$; i.e.\ violate our second
condition above.  Then, the theory is stringy only in the directions
of $x^{0,1}$.  A stringy spectrum in these directions can be removed
by the ghosts, such that the whole theory has a field theoretic
spectrum.  In fact, we have already seen that above, when we analyzed
the theory in the lightcone gauge.  However, if $G_{00} \sim \alpha'
\rightarrow 0$, while some of the other components of $G_{\mu\nu}$
remain of order one, then the contribution of the energy to the
dispersion relation $G^{00}p_0^2 \sim {1 \over \alpha'}$ is large, and
the transverse oscillator modes can be excited.  Therefore, in such a
scaling limit the transverse oscillators do not decouple and the
theory is stringy.

\newsec{New Noncritical String Theories}

In this section we are looking for a limit of the theory which focuses
on the nearly tensionless strings for $|E|\roughly< E_{cr}$ ($|\tilde
E|\roughly <1$).  Unlike the previous section, here we keep $\alpha'$
fixed, rather than scaling it to zero.  Hence the resulting theory is
a string theory and not a field theory.  The limit we look for is
space/time noncommutative and decouples from the bulk modes off the
branes including the gravitational modes of the theory.

Since we want the open string metric $G_{\mu\nu}$ to be fixed, we
should scale the closed string metric
\eqn\scaleg{\eqalign{
&-g_{00}=g_{11}=g \sim {1 \over 1-\tilde E^2}\cr
&|\tilde E|\to 1,}}
holding the other elements fixed (recall that we chose the closed
string metric to be diagonal).

Using \openparaa, the noncommutativity parameter $\theta$ can be
expressed in this limit as
\eqn\limitthe{\theta = 2\pi \alpha'G^{-1},}
and therefore it is finite.  Hence, in this limit the theory is
space/time noncommutative.

One of the surprising aspects of this theory is that the open strings,
which live on the branes, cannot turn into closed strings and
propagate into the bulk.  The energetics argument we gave around
equation \opencoupling\ prevents such processes.  We can also
understand it as follows.  For simplicity, let us rescale the open
string metric $G_{\mu\nu}$ to $\eta_{\mu\nu}$.  Then the closed string
metric $g_{\mu\nu}$ is equal to $\delta_{ij}$ for $i,j\not= 0,1$ and
$g_{00}=-g_{11}=-g=-1/(1-\tilde E^2)$.  Consider a process in which
open strings with energies $p^{(i)}_0$, spatial momenta
$(p^{(i)}_1,p^{(i)}_2,...)$ and masses $m^{(i)}$ attempt to escape
from the branes to the bulk by turning into closed strings.  The open
strings dispersion relations are of the form
$(p^{(i)}_0)^2=(p^{(i)}_1)^2+(p^{(i)}_2)^2+...+(m^{(i)})^2$.  As they
scatter to closed strings in the bulk, the total energy and momenta in
the directions of the branes are conserved but the metric used in the
dispersion relation is the closed string metric.  Therefore, for each
closed string we have ${1\over g} p_0^2 = {1 \over g} p_1^2 + p_2^2 +
... + m^2$.  Since $g=1/(1-\tilde E^2) \rightarrow \infty$ as $\tilde
E \to 1$, such a dispersion relation cannot be satisfied unless $p_0$
is infinite.  We see that we need to have an infinite amount of energy
in order to get off the branes -- the strings are confined to the
branes.

Since we scale $|E|\rightarrow E_{cr}$, the open string coupling
constant $G_s$ vanishes.  As a first attempt to have an interacting
theory, we scale $N$ to infinity holding the effective 'tHooft
coupling \thooftco
\eqn\thooftcoa{G_{eff}=NG_s=Ng_s\left(1-\tilde E^2\right)^\half}
fixed; i.e.\ we scale
\eqn\scalen{N \sim \left(1-\tilde E^2\right)^{-\half}.}

We conclude that our theory includes interacting open strings and the
underlying spacetime is noncommutative.  The open strings decouple
from the closed strings, and therefore also from the gravitational
sector of the theory.

It might appear surprising that we have open strings without closed
strings.  After all, closed strings should be found as poles in the
double twist diagram.  However, our limit with large $N$ suppresses
the nonorientable double twist diagram, while the orientable planar
diagrams survive.  Our theory might include composite closed strings
analogous to the QCD glueballs but these are not expected to be seen
in perturbation theory and should not be confused with the closed
strings which propagate off the branes.

If the theory confines and does produce its own glueballs, their
coupling constant should be proportional to $1/N$ by topological
considerations. Since we scale $N$ to infinity, the glueball coupling
constant vanishes. Therefore the theory is free in terms of these
variables. If a dual supergravity description exists in the strong
'tHooft coupling limit, we expect the glueballs to be described in
terms of classical supergravity in some dual bulk geometry.  Such a
description will be useful for $G_{eff}$ large.

Another possibility, is to keep $N$ fixed and scale $g_s$ to infinity
holding $G_{eff}$ fixed.  Thus we scale
\eqn\secondlimit{g_s\sim\left(1-\tilde E^2\right)^{-\half}.}
The strong coupling behavior of the theory can be analyzed
using string duality.  In the IIA theory we should use M-theory and in
the IIB theory we should use S-duality.

Let us focus on the case of the IIB theory.  We start with
\eqn\originallimi{\eqalign{
&g_s\sim (1-\tilde E^2)^{-\half} \cr
&g_{11}=-g_{00} \sim (1-\tilde E^2)^{-1} \cr
&g_{ij} \sim 1 \quad {\rm for} \quad i,j \not=0,1 .\cr}}
After an S-duality transformation we have
\eqn\aftersdu{\eqalign{
&g'_s\sim (1-\tilde E^2)^{\half} \cr
&g'_{11}=-g'_{00} \sim (1-\tilde E^2)^{-\half} \cr
&g'_{ij} \sim (1-\tilde E^2)^{\half} \quad {\rm for} \quad i,j
\not=0,1.\cr}}  
The lightest degrees of freedom in the bulk are the fundamental
strings which were originally D-strings.  Their coupling constant goes
to zero in the scaling limit.  The original fundamental strings are
D-strings after the duality transformation.  Although they are
strongly coupled, they are very heavy and are irrelevant for low
energy processes.  Of course, this last comment is consistent with the
previous discussion about the decoupling of these closed strings from
the modes on the branes. Since $g'_s \rightarrow 0$, we do not have
any closed string poles in the non-planar open string diagrams. The
only poles which could be generated are poles of closed D-strings,
which couple strongly. However, these are very heavy in the limit we
consider and cannot be put on shell for any finite energy of the
external open strings. Therefore, the energetics prevents such poles.

We also note that the nature of our D-branes changes.  If we started
with D3-branes, they remain D3-branes after the S-duality
transformation.  However, the background NS $B_{01}$ field becomes
background RR $B_{01}$ field.  So we end up with IIB theory at weak
coupling with D3-branes in background RR $B_{01}$ field.

We conclude that in the low energy limit we consider, with $g_s$
scaling to infinity as in \secondlimit, the theory includes
interacting open strings on the branes together with decoupled free
closed strings in the bulk.  The open string theory is decoupled from
gravity.  Again, the underlying spacetime on the D-branes is
noncommutative. In this limit $N$ is kept fixed and so are the
interactions of the QCD-like strings. We expect the theory to be dual
to a fully interacting theory of such strings.

Our theory is reminiscent of the little string theory of
\ref\lst{N.~Seiberg,
``New theories in six dimensions and matrix description of M-theory on
T**5 and T**5/Z(2),'' Phys.\ Lett.\ {\bf B408} (1997) 98
[hep-th/9705221].}.
In both cases we take a limit of string theory in which gravity
decouples to find a theory which does not appear to be a local quantum
field theory. 
\nref\coor{O.~Aharony, M.~Berkooz, D.~Kutasov and N.~Seiberg,
``Linear dilatons, NS5-branes and holography,''
JHEP {\bf 9810} (1998) 004 [hep-th/9808149].}%
\nref\minsei{S.~Minwalla and N.~Seiberg,
``Comments on the IIA NS5-brane,''
JHEP {\bf 9906} (1999) 007 [hep-th/9904142].}%
What are the observables of the theory?  In the little string theory,
the S-matrix elements of the underlying string theory lead to Green's
functions in the little string theory \refs{\coor,\minsei}.
Perhaps a similar construction exists for our theory but we have not
investigated this.

The thermodynamics of the theory is also interesting.  At weak
coupling, one expects the high temperature thermodynamics of the
theory to be that of a string theory with an exponentially growing
density of states and a Hagedorn transition. It would be interesting
to understand the behavior of the theory near the Hagedorn temperature
since this theory does not contain closed strings and decouples from
gravity. It would be interesting to understand the thermodynamics of
the theory at large $N$ with $G_{eff}$ held fixed and large, in
particular whether a Hagedorn spectrum still persists. The existence
of a possible supergravity dual may help towards this direction.

In \swnoncom\ the zero slope limit of D4-branes in a magnetic
background was related to D4-branes in a background electric field in
the limit as the electric field approaches the critical value.  These
two limits turned out to be related from an eleven dimensional
perspective in M-theory.  It would be interesting to see whether this
observation is relevant to our noncritical theory, as we use such a
limit.

\bigskip
\centerline{\bf Acknowledgements}

N.S. would like to thank the Theory Group at the University of Texas
for hospitality during part of this work.  We are greatful to
R. Gopakumar, S. Minwalla and A. Strominger for a number of
discussions concerning the various limits reported in this paper.  We
also thank M. Kleban, S. Shenker and E. Witten for useful discussions.
The work of N.S. was supported in part by DOE grant
\#DE-FG02-90ER40542. The work of L.S. and N.T. was supported in part
by NSF grant 980115.

\bigskip

\listrefs

\bye